\documentclass[11pt,twoside]{article}
\usepackage{atmp}
\usepackage{epsfig}
\usepackage{amsmath}
\usepackage{amssymb}
\copyrightnotice{2003}{7}{307}{330}       
\def\eqn#1#2{\begin{equation}#2\label{#1}\end{equation}}

\def\ha{{\frac{1}{2}}}

\def\IZ{\mathbb{Z}}

\def\im{\mbox{\,Im\,}}
\def\re{\mbox{\,Re\,}}
\def\betaH{\beta}
\def\betaI{\beta_I}
\def\TH{T_{\mbox{\scriptsize Hawking}}}
\def\Jmin{J^{\mbox{\scriptsize (min)}}}
\def\ddr{\frac{\partial}{\partial r}}
\def\warp{\left( 1 - \frac{1}{r} \right)}

\def\abs#1{\lvert #1 \rvert}
\def\nn#1#2{({\bf{#1}}, {\bf{#2}})}
\def\n#1{({\bf{#1}})}

\begin{document}

\vspace{-7mm}

\title{Asymptotic black hole\\ quasinormal frequencies}

\url{hep-th/0301173}

\vspace{-6mm}

\author{Lubo\v{s} Motl}  

\vspace{-9mm}

\address{Jefferson Physical Laboratory\\ Harvard University\\ Cambridge, MA 02138}
\addressemail{motl@feynman.harvard.edu}

\vspace{-6mm}

\author{Andrew Neitzke}

\vspace{-9mm}

\address{Jefferson Physical Laboratory\\ Harvard University\\ Cambridge, MA 02138}
\addressemail{neitzke@fas.harvard.edu}

\markboth{ASYMPTOTIC BLACK HOLE\ldots}{L. MOTL, A. NEITZKE}

\begin{abstract}
We give a new derivation of the quasinormal frequencies of
Schwarzschild black holes in $d \ge 4$ and 
Reissner-Nordstr\o{}m black holes in $d = 4$, in the limit of infinite damping.
For Schwarzschild in $d \ge 4$ we find that the asymptotic real part
is $\TH\log(3)$ for scalar perturbations and for some gravitational perturbations; this confirms a 
result previously obtained by other means in the case $d = 4$.  For 
Reissner-Nordstr\o{}m in $d=4$ we find a specific generally aperiodic behavior 
for the quasinormal frequencies, both for scalar perturbations and for 
electromagnetic-gravitational perturbations.  The formulae are obtained by 
studying the monodromy of the perturbation analytically continued to the complex plane;
the analysis depends essentially on the behavior of the potential in the ``unphysical'' region
near the black hole singularity.
\end{abstract}

\cutpage

\section{Introduction: why do we study\\ quasinormal modes?}

Recently it has been proposed \cite{dreyer} that the asymptotic behavior
of the high overtone black hole quasinormal frequencies captures important information
about the spectrum of black hole observables and about quantum gravity in
general.  More precisely, it has been suggested that the asymptotic real
part of the black hole
quasinormal frequencies coincides with the frequency emitted
by a black hole whose area decreases by an amount that is natural from the
point of view of discrete approaches to quantization of gravity such as
loop quantum gravity (LQG).  While such approaches
remain highly speculative, it is nevertheless interesting to understand
how far this argument can be pushed; moreover, even if this particular
line of inquiry has to be abandoned, the asymptotic spectrum of black hole
excitations is a fundamental property of the black hole which should 
eventually be good for something.  This paper is an attempt to simplify and
extend the calculations of the relevant ``experimental data.''

Let us briefly review the recent developments in loop quantum gravity which have led us to
investigate this asymptotic spectrum.  
(See \cite{rovelli,rovellitwo,thiemann} for 
increasingly lengthy reviews of 
LQG and \cite{rovellismolin,ashle} for the original derivations of the 
area quantization law in LQG.)  In LQG, the natural quantum $A_0$ of the horizon's area is proportional to the
so-called Barbero-Immirzi parameter $\gamma$ \cite{barbero,immirzi}, a parameter
in LQG which is arbitrary from the point of view of the microscopic definition
of the theory, but whose value
must be chosen in such a way that the black hole entropy comes out
correctly.  This comes about as follows:  one imagines that each quantum of
area $A_0$ carries a piece of discrete information.  It can be found in $k$
different microstates, and therefore it carries entropy $\log k$.
This must be equal to the known contribution of this area
quantum to the black hole entropy, namely\footnote{In this paper, we use
particle physics units with $c=\hbar=1$.} $A_0/4G$, so we find that
\eqn{areaquantum}{A_0 = 4G \log k.}
LQG predicts that $k=2\Jmin+1$, where $\Jmin$ is the minimal allowed 
spin carried by the spin networks. This means that
$k=2$ if the basic gauge group of LQG is taken to be $SU(2)$, or $k=3$ if 
the gauge group is $SO(3)$.  However, from the LQG point of view
there seems to be no reason to choose this strange number 
$\log(2)$ or $\log(3)$ (more precisely, $\log(2)/ \pi \sqrt 3$ or
$\log(3) / 2\pi\sqrt 2$)
for the Barbero-Immirzi parameter.

On the other hand, seemingly unrelated, one can consider 
the ``quasinormal modes'' of the black hole background. 
Quasinormal modes are perturbations of the background 
which are allowed to have complex frequency but whose boundary conditions 
must be ``purely outgoing'' both at the horizon {\it and} at infinity; we 
will review their precise definition in Section \ref{formu}.
This boundary condition singles out discrete complex frequencies $\omega_n$. As 
$n\to\infty$, the quasinormal frequencies of gravitational perturbations are 
known to behave as
\eqn{asympfreq}{\omega_n = \TH (2\pi i (n-1/2) + \log(3)).}
(Varying from the usual convention in general relativity, we parameterize the 
frequencies in terms of $\TH$, which is equal to $1/8\pi M G$ for a 
four-dimensional Schwarzschild black hole.) 
The imaginary part of \eqref{asympfreq} is not hard to understand.  The 
quasinormal modes determine the position of poles of a Green's function on 
the given background, and the Euclidean black hole solution converges to a 
thermal circle at infinity with the inverse temperature
$\betaH=1/\TH$; therefore it is not too surprising that the 
spacing of the poles in \eqref{asympfreq} coincides with the spacing $2\pi 
i \TH$ expected for a thermal Green's function. 

However, we are not aware of a similar classical explanation of
the term $\log(3)$ in \eqref{asympfreq}.  This term is remarkable because, as 
was first observed in \cite{dreyer}, it is precisely the same strange number
one needs to occur in the Barbero-Immirzi parameter of LQG if the gauge group
is taken to be SO(3).  There are also some heuristic arguments
which suggest that there might indeed be a reason to relate the two,
identifying the quasinormal mode with a process in which the black hole ejects
a single spin-network edge and thus reduces its area by $A_0$.

The result \eqref{asympfreq} was originally obtained by numerical calculations ten
years ago \cite{leaver,nollert,andersson,bachelot}.  At that time the term $\log(3)$ was
only known to an accuracy of six decimal places. In 1998, 
Hod conjectured \cite{hod} that this number was exactly $\log(3)$.  His 
conjecture was recently proven in \cite{lubosone}, using the methods of 
continued fractions initiated by Leaver \cite{leaver} and 
further refined by Nollert \cite{nollert}.

In this paper we would like to reproduce the results of \cite{lubosone},
using simpler functional methods which also allow us to extend the analysis 
to the cases of charged and higher-dimensional
black holes. 
We will find that the answer $\TH \log(3)$ is in fact universal for
Schwarzschild black holes in all dimensions, at least for scalar and some
gravitational perturbations.
For Reissner-Nordstr\o{}m
in $d=4$ we will find a more complicated generically aperiodic behavior,
both for scalar and electromagnetic-gravitational
perturbations.

Our computation depends on an analytic continuation
in the radial coordinate $r$, so it is worth mentioning
that analytic continuation in $r$ has been applied
to the determination of quasinormal modes many times before;
in particular, it was the basis of the numerical calculations 
in \cite{andersson} which confirmed that $\re \omega \to \TH \log(3)$ for Schwarzschild.  
A review of the use of complex coordinate methods
to compute black hole quasinormal frequencies has recently
appeared in \cite{glampand}, to which we refer the reader for
more extensive references.
We note that in our calculation of the leading-order asymptotic frequencies
no phase integral technology will be required; it will be
enough to use simply the first-order WKB approximation.

We believe that this puzzling behavior of the Green's functions
at large imaginary frequencies should contain some important information about black
holes and perhaps even quantum gravity.

\section{Quasinormal frequencies for $d=4$ Schwarzschild black holes} \label{formu}

\subsection{The differential equation}

In this section we will be concerned with 
properties of certain perturbations of a four-dimensional Schwarzschild geometry, described by
solutions of the differential equation \cite{reggewheeler}
\begin{equation} \label{de}
\left( - \left[ \left(1 - \frac{1}{r}\right) \ddr \right]^2 + V(r) - 
\omega^2 \right) \psi(r) = 0
\end{equation}
on the interval $1 < r < \infty$,
with the ``Regge-Wheeler'' potential
\begin{equation} \label{potential}
V(r) = \warp \left( \frac{l(l+1)}{r^2} + \frac{1 - j^2}{r^3} \right).
\end{equation}
The equation \eqref{de} describes the radial dependence of a spin-$j$ perturbation of the
background, with orbital angular momentum $l$ and time dependence $e^{-i \omega t}$.
We have set $2GM = 1$; in these units the horizon is at $r=1$.

The numerator $(1-j^2)$ which appears in \eqref{potential}
characterizes the spin $j$ of the perturbation for $j=0,1,2$, i.e. for perturbations
by scalar or electromagnetic test fields as well as perturbations of the 
metric itself.  (Actually, in the cases $j=1,2$, one has 
to expand in tensor spherical
harmonics and then \eqref{potential} is only correct if
we consider an ``axial'' perturbation; see e.g. \cite{zerilli}.)
It is nevertheless interesting to study
the problem for general $j$, because
our calculation relates this coefficient to Bessel's equation 
and shows that it is very natural to write the numerator in this form.
It might be useful to summarize its values for the
most important examples of $j$:
\eqn{spiny}{1-j^2 = \left\{
\begin{array}{rlrcl} 1:&\mbox{scalar perturbation}& j&=&0\\ 
0:&\mbox{electromagnetic perturbation}& j&=&1\\ 
-3:&\mbox{gravitational perturbation}& j&=&2 \end{array} \right.}

The factor $\warp$ which occurs in
\eqref{de}, \eqref{potential} is the ``warp factor'' in the Schwarzschild
geometry, which in Schwarzschild coordinates is simply the time component
$-g_{00}$ of the metric.

We wish to study \eqref{de} on the physical region $1 < r < \infty$ and we begin
by describing the asymptotic solutions.  To simplify the differential
operator which appears it is convenient to 
define the ``tortoise coordinate'' $x$ with the property that
\begin{equation} \label{tortoise-origin}
dx = \warp^{-1} dr.
\end{equation}
Integrating \eqref{tortoise-origin} gives
\begin{equation} \label{tortoise}
x = r + \log(r-1).
\end{equation}
Then $x$ takes values $-\infty < x < \infty$, and \eqref{de} becomes
\begin{equation} \label{dex}
\left( - \frac{\partial^2}{\partial x^2} + V[r(x)] - \omega^2 \right) \psi = 0.
\end{equation}
Since $V[r(x)] \to 0$ as $x \to \pm \infty$,
the solutions behave as $\psi \sim e^{\pm i \omega x}$ at infinity.

\subsection{Analytic continuation}

In our analysis we will find that it is essential to extend \eqref{de} beyond the
usual physical region $1 < r < \infty$.
Continued to the whole complex $r$-plane, \eqref{de} is an
ordinary differential equation with regular singular points at $r=0$, $r=1$ and an irregular singular point
at $r=\infty$; and
by the general theory of differential equations, any solution of \eqref{de} in the physical
region extends to a solution on the $r$-plane.  However, this solution
may be multivalued around the singular points $r=0$, $r=1$.  This multivaluedness will play
a crucial role in our analysis, which is essentially concerned with the computation of the
monodromy around a particular closed contour in the $r$-plane.  For convenience, to avoid having 
to deal with multivalued functions, we can put branch cuts in the $r$-plane from $r=0$ and $r=1$ 
and require that $\psi(r)$ satisfy \eqref{de} everywhere except at the cuts.  In this approach the monodromy
is defined by the discontinuity across the cut.

\begin{minipage}{60mm}
\epsfig{file=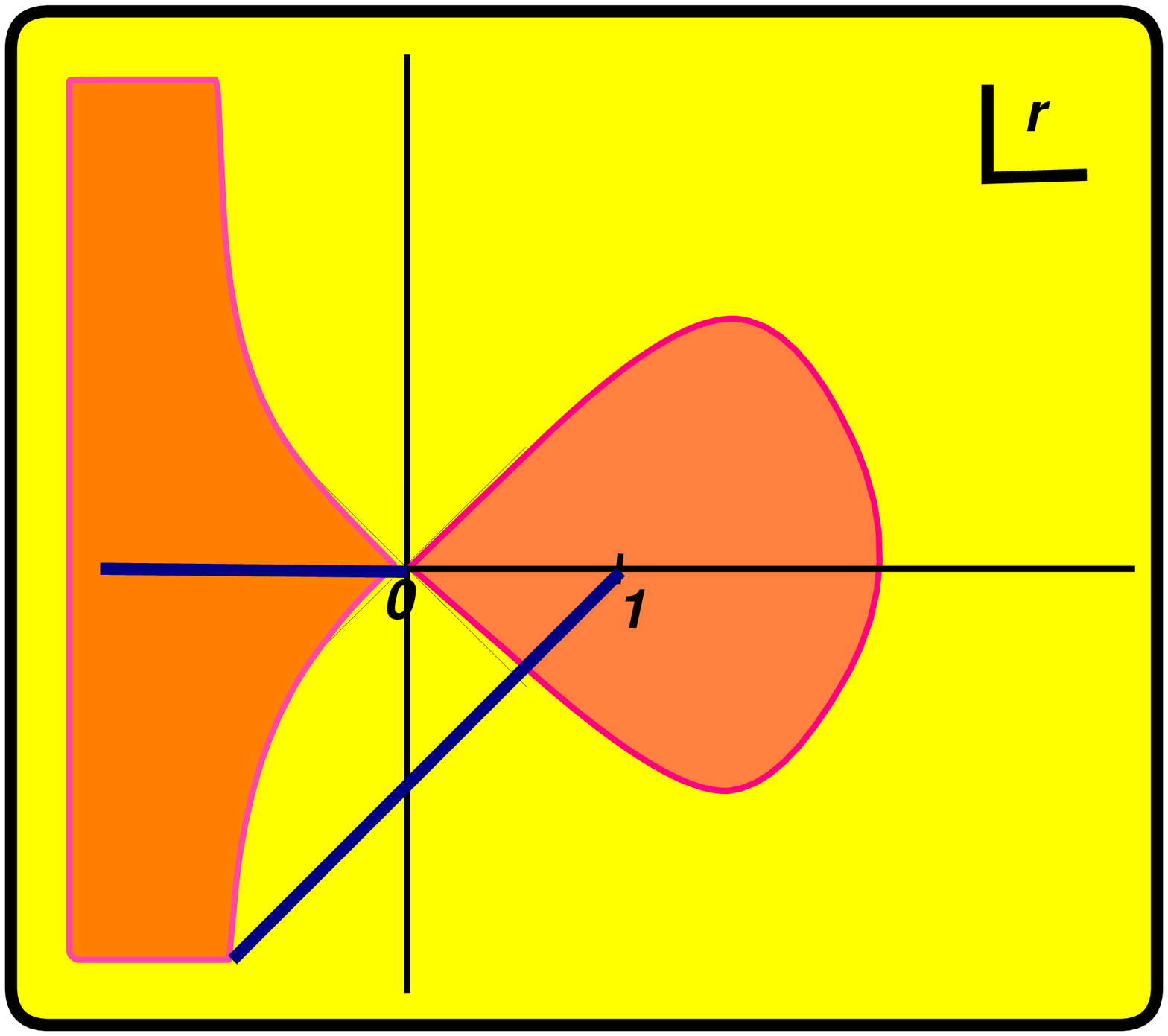,width=55mm}
\end{minipage}
\begin{minipage}{55mm}
{\small {\bf Figure 1:} The $r$-plane, with regions where $\re x<0$ denoted by
a darker color, and a convenient choice of branch cuts shown.}
\end{minipage}

\vspace{5mm}

We will find it very convenient to exploit the simpler representation \eqref{dex} for the differential
equation.  After passing to complex $r$ this introduces an extra complication, since the relation 
\eqref{tortoise} between $x$ and $r$ is itself multivalued because of the term $\log (r-1)$.  As a result, as 
we travel around a contour in the $r$-plane, the corresponding contour in the $x$-plane may not be closed.
In the language of branch cuts we would say that $x$ jumps when $r$ crosses the branch cut which
emanates from $r=1$.

\vspace{5mm}

Although $x$ is not uniquely defined as a function of $r$, the ambiguity is only in the imaginary
part, so $\re x$ still makes sense.  It will be important for us to know the sign of $\re x$ in the
various regions of the $r$-plane; the structure is as shown in Figure 1.  We also show
a convenient choice of branch cuts (to economize, one should take the branch cut to 
be the same for $\psi(r)$ and for $\log(r-1)$) although nothing we say will depend on this choice.

\subsection{Definition of quasinormal modes}

Because the potential $V(r)$ is positive and decays to zero at both ends, 
there are no discrete normalizable bound states.
Nevertheless we may look for discrete {\it quasinormal} states, 
counterparts of quasistationary states in quantum
mechanics because their frequency is also allowed to be complex.  
The mathematical theory of these quasinormal modes is more
intricate than for ordinary normal modes; see e.g. \cite{kokkotas} for
a more precise discussion than we will give here, as well as
an excellent review of the substantial literature on quasinormal modes
of black holes.

\vspace{4mm}

Roughly, a quasinormal mode with frequency $\omega$
is supposed to be a solution of \eqref{dex} where we require ``outgoing''
behavior at both ends of the physical region:
\begin{align} \label{naive1}
\psi & \sim e^{-i \omega x} \text{\quad as\quad} x \to +\infty, \\
\psi & \sim e^{+i \omega x} \text{\quad as\quad} x \to -\infty. \label{naive2}
\end{align}
Since the differential operator in \eqref{de} would be self-adjoint 
acting on normalizable wavefunctions, and hence would have to have real eigenvalues,
there are no quasinormal modes with $\im \omega < 0$.  
So we take $\im \omega > 0$, and then in \eqref{naive1}, \eqref{naive2} we are asking for the exponentially
decreasing component to be absent at both ends.  It is actually quite delicate to define
exactly what this means!  If $V[r(x)]$ were strictly vanishing at $x \to \pm \infty$ then one 
could just require that $\psi$ be strictly proportional to $e^{\pm i \omega x}$ as appropriate, 
but for more general $V[r(x)]$ we do not know any way of defining the boundary condition
using only the behavior of $\psi$ on the real line.  This problem was also discussed in \cite{ns} where
it was suggested that the appropriate condition would be obtained by finding an analytic extension
of $\psi$ to $\im \omega<0$ and demanding exponential \textit{decrease} there; this was further 
shown to be equivalent to looking for poles in an appropriate Green's function.  In \cite{leaver} another 
criterion was given in terms of the coefficients of a power series representation for $\psi$; this
criterion is what was used in the analytical calculation of the asymptotic quasinormal 
frequencies in \cite{lubosone}.

Here we wish to use another way of defining the boundary
condition, by analytic continuation in $r$ instead of
$\omega$.  (The method we describe below has been studied before, e.g. in \cite{bd-horizon};
see also \cite{glampand} for a recent discussion and additional references.)
By considering complex $r$ we can easily define 
the boundary condition at the regular singular point $r=1$, because the functions 
$e^{\pm i \omega x}$ are distinguished by their monodromy there.  Namely, 
$e^{\pm i \omega x} = (r-1)^{\pm i \omega} e^{\pm i \omega r}$,
so to get $\psi \sim e^{i \omega x}$ near $r=1$ we simply require 
that $\psi(r)$ have monodromy $e^{- 2 \pi \omega}$ on a counterclockwise circle around $r=1$.
What is more subtle is to define the boundary condition at $r=\infty$, but this can be done by
analytically continuing $\psi(r)$ via ``Wick rotation'' to the line
$\im \omega x = 0$, where the asymptotic behavior of the solutions is purely
oscillatory and hence one can really pick out a particular solution by specifying
its asymptotics.
In the limit where $\omega$ is almost pure imaginary, the line 
$\im \omega x = 0$ is slightly tilted off the line $\re x = 0$. 
There are two possible directions for the Wick
rotation and we choose the rotation through an angle smaller than $\pi/2$.
For the time being we assume $\re \omega > 0$; in this case $x = \infty$ gets
rotated to $\omega x = \infty$, so our boundary condition ``at $r = \infty$'' is actually 
\begin{equation} \label{bd-infty}
\psi(r) \sim e^{-i \omega x} \textrm{\quad as\quad} \omega x \to \infty.
\end{equation}

\subsection{Computation of quasinormal frequencies}

Having defined the quasinormal modes we now proceed to study their frequencies, in the
limit $\im \omega \gg \re \omega$.
Indeed, until the end of this computation we will behave as if $\omega$ were purely imaginary.

We will compute in two different ways 
the monodromy of $\psi(r)$ around the contour shown in Figure 2, beginning and ending
at the point marked A.

\epsfig{file=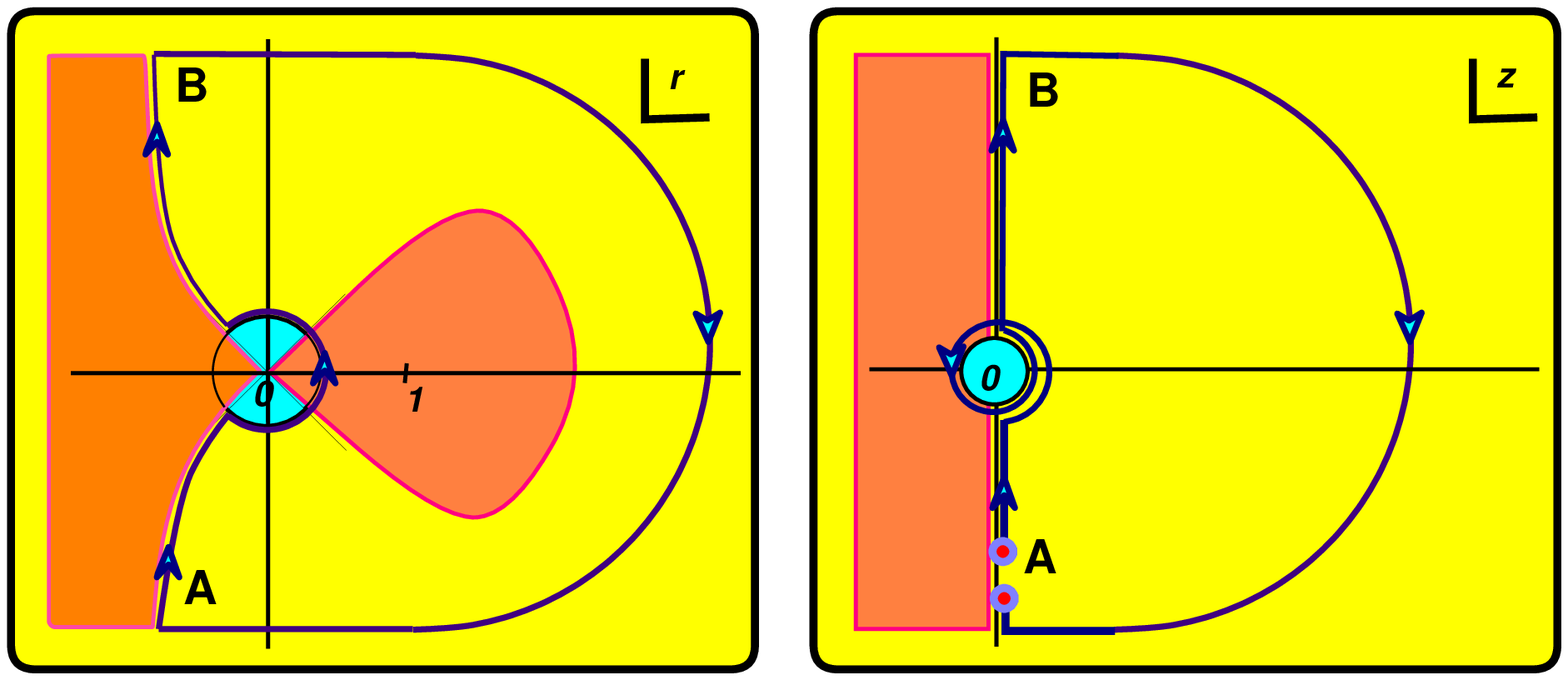,width=114mm}

\vspace{-6mm}

\begin{quotation}
{\small {\bf Figure 2:} Contour for the 
calculation of quasinormal frequencies, shown in
terms of $r$ and in terms of $z = x-i \pi$.  The 
singularities at $r = 0$ (equivalently $z = 0$) 
and $r = 1$ (equivalently $z=-\infty$)
are marked.  The rotation by $3 \pi / 2$ in the $r$ plane 
becomes a rotation by
$3 \pi$ in the $z$ plane as follows from \eqref{double-cover}. 
The monodromy of $\log(r-1)$ around $r=1$ causes a $2\pi i$ discontinuity in the 
path in the $z$-plane. The regions with $\re x <0$ are denoted by
a darker color. The wave function is well approximated by a sum of Bessel
functions in the blue circle around $r=0$.}
\end{quotation}

\vspace{-3mm}

It will prove convenient to use the coordinate $z = x-i \pi$ instead of $x$.  Actually, this definition
is slightly imprecise since we have not specified which branch of the logarithm in $x(r)$ we should choose.
Depending on the branch, the contour in the $z$-plane will approach some point $z = 2 \pi i n$ as $r$ approaches $0$;
for convenience we choose the branch so that $n = 0$.

First we compute the monodromy by matching the asymptotics along the line $\im \omega z = 0$. 
Beginning at A, 
the plane wave behavior \eqref{bd-infty} can be extrapolated toward the interior 
since the term $\omega^2$ dominates $V$ in \eqref{dex} away from a small neighborhood around $r = 0$.  As we
extrapolate up from A we reach $r = 0$ at $z = 0$, and near this point we have
\begin{equation} \label{double-cover}
z = r + \log(r-1) - i \pi \sim -\ha r^2,
\end{equation}
so in a neighborhood of $r = 0$ we may write
\begin{equation} \label{potential-sing}
V[r(z)] \sim \frac{j^2-1}{4z^2}.
\end{equation}
Thus \eqref{de} may be simplified near $r=0$ to
\begin{equation} \label{bessel-ode}
\left( -\frac{\partial^2}{\partial z^2} 
+ \frac{j^2 - 1}{4z^2} - \omega^2 \right) \psi(z) = 0.
\end{equation}
This equation can be exactly solved in terms of Bessel 
functions; for generic $j$, the general solution is
\begin{equation} \label{bessel-sol}
\psi(z) = A_+ c_+\sqrt{\omega z} J_{j/2} (\omega z) + A_- 
c_-\sqrt{\omega z} 
J_{-j/2}(\omega 
z).
\end{equation}
We choose the normalization factors $c_+,c_-$
so that the asymptotics are 
\begin{equation} \label{bessel-asymp-before}
c_\pm
\sqrt{\omega z} J_{\pm j/2} (\omega z) \sim 2\cos (\omega z - 
\alpha_{\pm}) 
\textrm{\quad as\quad} \omega z \to \infty,
\end{equation}
where we defined
\begin{equation}
\alpha_{\pm} = \frac{\pi}{4} (1 \pm j).
\end{equation}
The phase shifts by $\alpha_\pm$ will be very important for our analysis.
Using \eqref{bessel-sol} and \eqref{bessel-asymp-before} 
together with the boundary condition \eqref{bd-infty}
which states that $e^{i \omega z}$ must be absent as $\omega z \to \infty$, we have
\begin{equation}
A_+ e^{-i \alpha_+} + A_- e^{-i \alpha_-} = 0,
\end{equation}
while for the coefficient of $e^{-i \omega z}$ we have
\begin{equation}
\psi(z) \sim \left( A_+ e^{i \alpha_+} + A_- e^{i \alpha_-} \right) e^{-i 
\omega z} \textrm{\quad as\quad} \omega z \to \infty.
\end{equation}
To follow the contour through to B we now have to turn through an angle $3 \pi / 2$ around $r = 0$, or equivalently
$3 \pi$ around $z = 0$.  Using the fact that 
\begin{equation}
J_{\pm j/2}(\omega z) = z^{\pm j/2} \phi(z),
\end{equation}
where $\phi$ is an even holomorphic function of $z$,
after the $3 \pi$ rotation the asymptotics are
\begin{equation} \label{bessel-asymp-after}
c_\pm \sqrt{\omega
z} J_{\pm j/2} (\omega z) \sim e^{6i \alpha_\pm} 2\cos (- \omega z - 
\alpha_{\pm}) \textrm{\quad as\quad} \omega z \to -\infty.
\end{equation}
Then using \eqref{bessel-sol}, \eqref{bessel-asymp-after} we get 
the asymptotics at B,
\begin{equation}
\psi(z) \sim \left(A_+ e^{5 i \alpha_+} + A_- e^{5 i \alpha_-} \right) 
e^{-i \omega z} + \left(A_+ e^{7 i \alpha_+} + A_- e^{7 i \alpha_-} \right) 
e^{i \omega z} \end{equation}
as $\omega z\to-\infty$.
Finally, we can continue along the large semicircle in the right half-plane from B back to A.  Since we are in the region where $\omega^2$ dominates the
potential in \eqref{dex} 
we essentially just have plane wave behavior; this approximation can be trusted at least for the coefficient of $e^{-i \omega z}$,
which therefore remains unchanged as we continue back to A and complete the contour
(however, the same cannot be said for the coefficient of $e^{i \omega z}$, which makes only an exponentially small contribution to $\psi(z)$ in
the right half-plane and hence can be modified by the small correction terms in the WKB approximation.)  In the end we find that
the monodromy around the contour multiplies the coefficient of $e^{-i \omega z}$ by a factor
\begin{equation} \label{monodromy}
\frac{A_+ e^{5i\alpha_+} + A_-e^{5i\alpha_-}}{A_+ e^{i\alpha_+} + A_-e^{i\alpha_-}} = \frac{e^{6i\alpha_+} - e^{6i\alpha_-}}{e^{2i\alpha_+}-e^{2i\alpha_-}} = -\frac{\sin 3\pi j/2}{\sin \pi j/2} = -(1 + 2 \cos \pi j).
\end{equation}

On the other hand, we can also obtain the monodromy action by observing that the only singularity of $\psi(r)$ or $e^{-i \omega z}$ 
inside the contour is at $r=1$.  By our boundary condition going around $r=1$ clockwise just multiplies $\psi(r)$ by $e^{2 \pi \omega}$,
but it multiplies $e^{-i \omega z}$ by $e^{-2 \pi \omega}$, so the coefficient of $e^{-i \omega z}$ in the asymptotics of 
$\psi(r)$ must be multiplied by $e^{4 \pi \omega}$ after the full round trip.  Comparing with \eqref{monodromy} gives
\begin{equation} \label{answer}
e^{4 \pi \omega} = -(1 + 2 \cos \pi j).
\end{equation}

Note that if we had chosen $\re \omega < 0$ we would have had to Wick rotate
in the opposite direction, hence set our boundary conditions at B instead of A, and run the contour in the opposite direction; hence we would have
gotten \eqref{answer} with $e^{-4 \pi \omega}$ instead of $e^{4 \pi \omega}$.  This is as expected since if $\omega$ is a quasinormal frequency
then $-\omega^*$ must also be one; the corresponding wave functions
are simply complex conjugates of one another.
The two possibilities can be summarized by the equation
\eqn{unified}{\exp\left(\epsilon\left(\!\re \omega\right) 4 \pi \omega\right) = -(1 + 
2 \cos \pi 
j),}
implying
\begin{equation} \label{unif-simple}
\abs{4 \pi \re \omega} = \log\abs{1+2\cos \pi j}.
\end{equation}
In the case $j = 0,2$ we recover the numerical result
of \cite{leaver, nollert, andersson, bachelot} 
which was so important for \cite{dreyer, hod}:
\begin{equation} 
4 \pi \omega = (2n + 1) \pi i \pm \log(3).
\end{equation}
If our analysis of the dependence on $\epsilon(\!\re \omega)$ is correct,
it implies $\abs{1+2\cos \pi j} \ge 1$.  
Hence for $j\in (1/2,3/2)+ 2 \IZ$, there cannot be any 
asymptotic quasinormal modes, with the possible exception of the case $j\in 
2\IZ+1$.  (This case is exceptional in our analysis for another reason as
well:  if $j=1$, then our form \eqref{potential-sing} for the leading singularity
vanishes.  A more careful analysis would be desirable to check whether 
our results still hold in this case.)

We should mention that \eqref{unified} is equivalent to a more complicated 
equation derived in \cite{lubosone}, namely
\eqn{unifiedlubosone}{\tan(\pi i \omega)
\tan(\pi i \omega + \pi j / 2)
\tan(\pi i \omega - \pi j /2) = i\epsilon(\!\re \omega)}
This equivalence can be shown by multiplying 
\eqref{unifiedlubosone} by 
\eqn{coscoscos}{\cos (\pi i \omega)\cos(\pi i \omega + \pi j / 
2)\cos(\pi i \omega - \pi j /2)}
and expressing the sines and cosines in terms of exponentials.

\subsection{Some convenient notation} \label{notation}

For later convenience we now describe the preceding computation in a more compact notation.  First write
\begin{equation}
\n{m}_A = (A_+ e^{i m \alpha_+} + A_- e^{i m \alpha_-})
\end{equation}
and
\begin{equation}
\nn{m}{n}_A = \n{m}_A e^{-i \omega z} + \n{n}_A e^{i \omega z}.
\end{equation}

Then the asymptotic matching
through the Bessel region which we did by comparing \eqref{bessel-asymp-before} and 
\eqref{bessel-asymp-after}
shows that a 3$\pi$ clockwise rotation in $z$ transforms the asymptotic
behavior by
\begin{equation} \label{mono3pi}
3 \pi: \nn{m}{n} \to \nn{n+6}{m+6},
\end{equation}
where the swap between $m$ and $n$ arises because of the swap $\omega z \leftrightarrow -\omega z$ needed to 
compare \eqref{bessel-asymp-before} and \eqref{bessel-asymp-after}.

In the case of four-dimensional Schwarzschild we read off the 
initial asymptotics from \eqref{bessel-asymp-before} as $\psi \sim \nn{1}{-1}_A$.  
Since the coefficient of $e^{i \omega x}$ had to vanish at A we got the constraint 
\begin{equation} \label{constraint-bd}
\n{-1}_A = 0.
\end{equation}
The $3\pi$ rotation \eqref{mono3pi} gave 
\begin{equation}
\nn{1}{-1}_A \to \nn{5}{7}_A,
\end{equation}
so following the coefficient
of $e^{-i \omega x}$ back around the semicircle gives its monodromy as $\n{5}_A / \n{1}_A$,
which we computed using the constraint \eqref{constraint-bd}, obtaining
\begin{equation}
\frac{\n{5}_A}{\n{1}_A} = -(1+2\cos \pi j).
\end{equation}

\section{Quasinormal frequencies for other black holes}

Having reproduced the result of \cite{lubosone} for 
$d=4$ Schwarzschild, we now apply the same technology to two other classes of black hole.  The 
generalizations are surprisingly straightforward.

\subsection{Higher-dimensional Schwarzschild black holes}

First we consider Schwarzschild black holes in arbitrary $d \ge 4$.
We will verify a
conjecture from \cite{lubosone}:  namely, the asymptotic quasinormal 
frequencies of the Schwarzschild black hole with $\re \omega > 0$
satisfy
\eqn{higherf}{e^{\betaH\omega} = -3}
where $\betaH$ is the Hawking temperature of the black hole.
(Strictly speaking, we will prove this only for scalar
perturbations and for certain tensor modes; 
we do not know the appropriate potentials for general perturbations 
in $d>4$.)

To write the generalization of
\eqref{de} to $d \ge 4$ we need the warp
factor for $d$-dimensional Schwarzschild,
\eqn{warpd}{-g_{00} = f(r) = 1-\frac{1}{r^{d-3}},}
where we again chose units in which the horizon is at $r=1$.
Then the perturbation satisfies
\begin{equation}\label{de-hs}
\left[ - \left( f(r) \ddr \right)^2 + V(r) - \omega^2 \right] \psi(r) = 0,
\end{equation}
where $V(r)$ is known for scalar perturbations to be \cite{cardoso}
\begin{equation} \label{potential-hs}
V(r) = f(r) \left[ \frac{l(l+d-3)}{r^2} + \frac{(d-2)(d-4)f(r)}{4r^2} + \frac{(d-2)f'(r)}{2r} \right].
\end{equation}
As in the $d=4$ case, this $V(r)$ grows large only at $r=0$, and direct
computation gives the coefficient of the singularity as
\begin{equation} \label{leadingpot-hs}
V(r) \sim -\frac{(d-2)^2}{4r^{2d-4}}.
\end{equation}

\begin{minipage}{60mm}
\epsfig{file=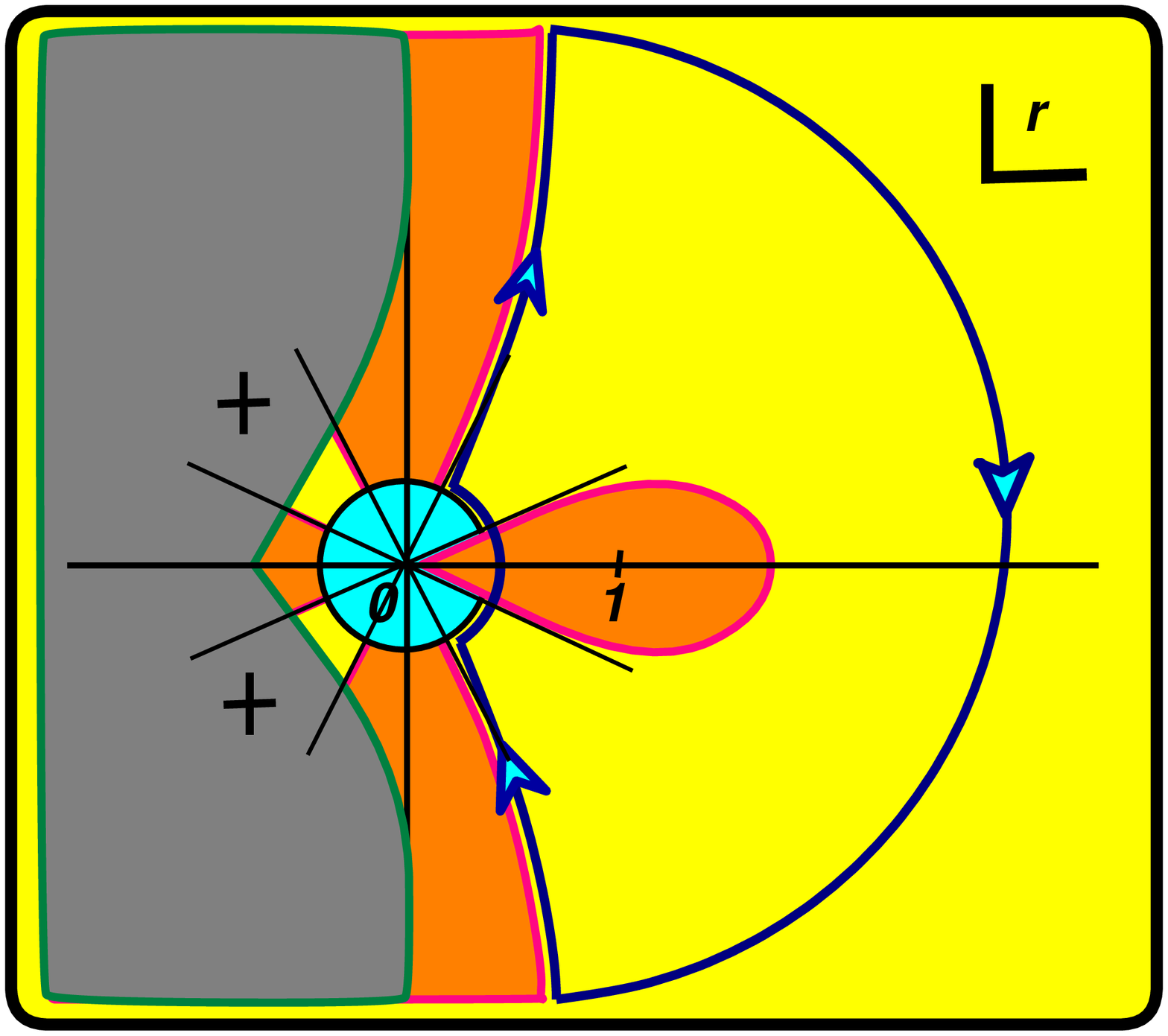,width=55mm}
\end{minipage}
\begin{minipage}{55mm}
{\small {\bf Figure 3:} Regions with $\re z<0$ (darker) in the
$r$-plane for a six-dimensional Schwarzschild black hole, together
with a contour for the calculation of quasinormal frequencies.  The left
part of the $r$-plane is shown in shadow because
$\re z$ can only be uniquely defined in the right part; the extra
fictitious horizons are marked by crosses.}
\end{minipage}

Now we want to describe the tortoise coordinate in $d$ spacetime 
dimensions.  From the warp factor \eqref{warpd} we get the requirement
\eqn{tortd}{dz = \left( 1 - \frac{1}{r^{d-3}} \right)^{-1} dr,}
which can be integrated to give
\eqn{tortdd}{z = r + \sum_{j=0}^{d-4}
\frac{e^{2\pi i j/(d-3)}}{d-3}
\log(1-r e^{-2\pi i j/(d-3)})
}
where we chose the additive constant so that $z=0$ for $r=0$.  From \eqref{tortdd}
we obtain the behavior of $\re z$ as shown in Figure 3.  Unlike the 
situation in $d=4$, \eqref{tortdd} includes phases multiplying the logarithms which
come from the fictitious ``horizons'' $r = e^{2 \pi i j/(d-3)}$, so that $\re z$ is
not well defined for arbitrary $r$; but so long as we stay in the right part of the 
$r$-plane it is well defined.

Near $r=0$, $d-2$ 
terms in the Taylor expansion of \eqref{tortdd} near $r=0$ cancel and
\eqn{rzerod}{z \sim - \frac{r^{d-2}}{d-2}, }
generalizing the behavior \eqref{double-cover} in the $d=4$ case.
Actually, this cancellation can be seen directly from \eqref{tortd}, by neglecting the term
$1$ on the right hand side.

Now we can express the singularity of the potential in terms of $z$; combining
\eqref{leadingpot-hs} and \eqref{rzerod} we find simply
\begin{equation} \label{leading-z}
V[r(z)] \sim -\frac{1}{4z^2}
\end{equation}
Note that this is exactly the same as the behavior for scalar perturbations in $d=4$;
namely, it agrees with \eqref{potential-sing} at $j=0$.  The other case for which we know the
potential is a particular family of tensor metric perturbations studied in \cite{gibbons}, 
which exist only in $d>4$; for these the leading singularity corresponds again to $j=0$.
(One might have expected $j=2$, but in our final result $j$ appears only in $\cos \pi j$,
so that $j=0$ and $j=2$ are indistinguishable.)
We conjecture that with some
suitable interpretation this form 
for the potential will actually persist for all $j$, so that the $-1$ may be
replaced by $j^2-1$ in \eqref{leading-z}, but the skeptical reader may just
set $j=0$ in the rest of this section.

At any rate, writing the potential in this form
the Bessel function behavior of the solutions near $r=0$, expressed in terms of $z$, 
is just as it was in $d=4$, and again we have plane wave behavior away from a
neighborhood of $r=0$.  
Furthermore, although the left part of the $r$-plane is 
rather complicated, the 
behavior of $\re x$ on the right part is essentially unchanged from $d=4$.  So we 
can repeat the argument from the previous section.  The 
calculation of the monodromy by matching asymptotics in the $z$-plane 
works exactly as in four 
dimensions, while the factor $e^{4 \pi \omega}$ from the monodromy around $r=1$ 
is modified to $e^{4 \pi \omega / (d-3)}$, because of the
coefficient $1/(d-3)$ multiplying
$\log(1-r)$ in \eqref{tortdd}.

Fortunately, this same factor appears in the formula for the Hawking temperature
of a $d$-dimensional black hole.  Namely, recall that the Hawking temperature
can be computed as the surface gravity of the black hole:
\eqn{surfg}{\TH = \frac{f'(1)}{4\pi}
= \frac{d-3}{4\pi},}
giving $\betaH = 4\pi / (d-3)$.  Hence the monodromy of the coefficient of 
$e^{-i \omega x}$  around the contour
is $e^{\betaH \omega}$, and comparing this to the result from Bessel asymptotics gives
\begin{equation}
e^{\betaH \omega} = -(1 + 2\cos \pi j),
\end{equation}
as desired.

In fact the agreement between the surface gravity and the coefficient of the logarithm
in the tortoise coordinate is not a coincidence and holds more generally.  
This is because a horizon is characterized by the condition $f(r_H) = 0$,
and the tortoise coordinate near $r_H$ then obeys
\begin{equation}
z = \int dz = \int \frac{dr}{f(r)} \approx \int \frac{dr}{(r-r_H) f'(r_H)} = \frac{1}{f'(r_H)} \log (r-r_H) 
\end{equation}
while the surface gravity is $f'(r_H) / 4\pi$.

Our result implies that for scalar and some gravitational perturbations
\begin{equation} \label{schw-result}
\re \omega = \pm \TH \log(3).
\end{equation}
In the revised version of \cite{kunstatter}, considerations of Bohr-Sommerfeld quantization (as
originally proposed by Bekenstein and amplified e.g. in \cite{bek2}) led to the conjecture 
that $\re \omega = \pm \TH \log(k)$ for some $k$; our calculation establishes this conjecture
and furthermore determines the integer $k=3$.

\subsection{Four-dimensional non-extremal Reissner-Nordstr\o{}m black holes}

Finally we want to extend our results to non-extremal Reissner-Nordstr\o{}m black holes
in $d=4$.  In this case the warp factor is 
\begin{equation}
f(r) = 1 - \frac{2GM}{r} + \frac{Q^2}{r^2} = \frac{(r-1)(r-k)}{r^2}
\end{equation}
where we have fixed our units by setting 
$GM = (1+k)/2$, $Q^2 = k$, so that the outer horizon is at $r=1$ and 
the inner horizon is at $r = k \le 1$.  The tortoise coordinate is therefore determined by
\begin{equation}
dz = dr \frac{r^2}{(r-1)(r-k)},
\end{equation}
which integrates to
\begin{equation} \label{rn-tortoise}
z = r + \frac{\log(1-r) - k^2\log(1-r/k)}{1-k}
\end{equation}
where we have again fixed the constant so that $z = 0$ at $r = 0$.  Then near $r =0$ we find
\begin{equation} \label{zclose-rn}
z \sim \frac{r^3}{3k}.
\end{equation}
The form of the potential can be found in \cite{berti2}:  
both for scalar perturbations and for axial electromagnetic-gravitational perturbations (which are coupled in this
background, so we cannot talk about electromagnetic or gravitational perturbations
separately) the potential is singular only near $r = 0$, and the leading singularity is
\begin{equation} \label{leading-potl-rn}
V(r) \sim \left\{ \begin{array}{cl}        
              -2k^2 / r^6 & \textrm{scalar perturbation} \\
              +4k^2 / r^6 & \textrm{electromagnetic-gravitational perturbation}
              \end{array} \right.
\end{equation}
Using \eqref{zclose-rn} we can rewrite this in the familiar form
\begin{equation}
V[r(z)] \sim \frac{j^2 - 1}{4 z^2}
\end{equation}
(with all the $k$ dependence disappearing)
and, remarkably, the values of $j$ which come out are nice:
\begin{equation} \label{j-rn}
j = \left\{ \begin{array}{cl}
              1/3  & \textrm{scalar perturbation} \\
              5/3  & \textrm{electromagnetic-gravitational perturbation}
            \end{array} \right.
\end{equation}

\epsfig{file=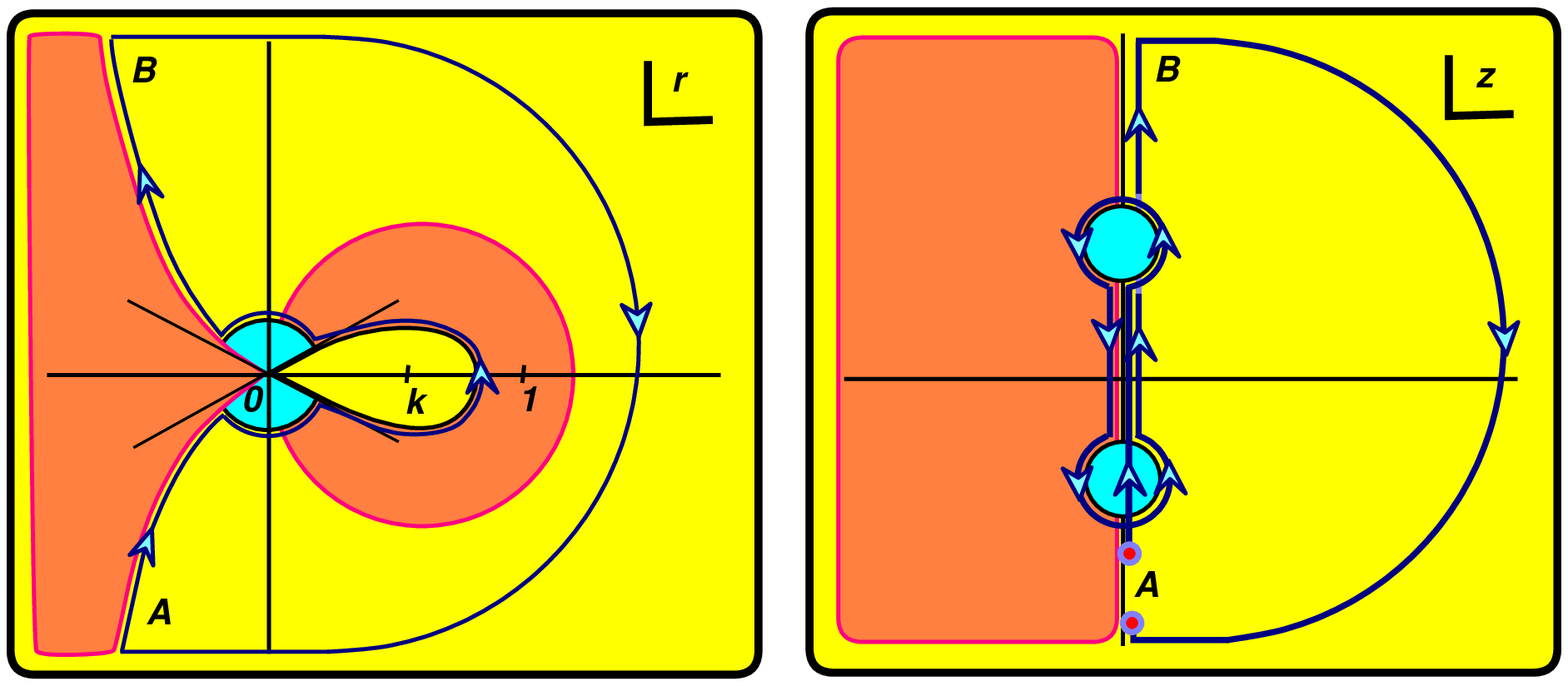,width=114mm}

\begin{quotation}
{\small {\bf Figure 4:} Regions with $\re(z)<0$ (darker) in the
$r$-plane for four-dimensional non-extremal Reissner-Nordstr\o{}m, 
with a contour for the calculation of quasinormal frequencies, shown both in the 
$r$-plane and the $z$-plane.}
\end{quotation}

Hence the Bessel function behavior near $r = 0$ will be unchanged from the previous examples,
and our calculation will proceed in an almost identical manner.
However, in this case we must use a slightly more complicated contour, going around a ``lobe'' as shown in Figure~4; 
the point is that we want to capture only the outer horizon at $r=1$, where
we know the boundary conditions and hence know the monodromy, and avoid the inner horizon at $r=k$.
As a result this computation is only valid in the non-extremal case $k < 1$.  

The corresponding
picture in the $z$-plane looks rather complicated because a given $z$ may or may not lie in a ``Bessel region''
depending on whether or not the $r$-plane contour is passing near $r=0$ 
(the potential is a multivalued function of $z$ but an ordinary function of $r$.)  In Figure~4
this is indicated by letting the contour pass over or under the blue circles when $z$ is \textit{not} passing
through a Bessel region; it is easiest to understand that figure by redrawing it yourself, following the 
given $r$-plane contour starting at A and tracing out the corresponding $z$-plane behavior.

The matching of asymptotic expansions in this case can be done following the recipe given in
Section \ref{notation}.  As before, we begin at the point marked A with the asymptotic behavior $\psi \sim \nn{1}{-1}_A$,
and we have the boundary condition
\begin{equation} \label{c1}
\n{-1}_A = 0.
\end{equation}
Going through a $2\pi$ rotation in $z$ gives (compare \eqref{mono3pi})
\begin{equation} \label{mono2pi}
2\pi: \nn{m}{n} \to \nn{m+4}{n+4}
\end{equation}
which is in this case $\nn{1}{-1}_A \to \nn{5}{3}_A$.  Then we can go out along the
lobe where in our approximation the behavior is purely oscillatory.  
After going around the lobe 
we have to return to the Bessel region a second time.
However, the asymptotics in terms of $z$ will be different this time:  we chose the branch of $z$ so
that the first time we entered the Bessel region it was at $z=0$, but now we have traveled an additional
distance in $z$ (from changing branches of the second logarithm in \eqref{rn-tortoise}), namely
\begin{equation}
\delta = - 2 \pi i k^2 / (1-k).
\end{equation}
So we may write the solution as
\begin{equation} \label{phased-asymp}
\psi(z) = B_+ c_+\sqrt{\omega (z-\delta)} J_{j/2} (\omega (z-\delta)) + B_- c_-\sqrt{\omega (z-\delta)} J_{-j/2}(\omega (z-\delta)).
\end{equation}
Introducing a slight modification of our previous notation, 
we write
\begin{equation}
\n{n_\delta} = e^{i \omega \delta} \cdot \n{n}.
\end{equation}
Then from \eqref{phased-asymp} we obtain $\psi \sim \nn{-1_\delta}{1_{-\delta}}_B$,
giving the constraint
\begin{equation} \label{c2}
\nn{5}{3}_A = \nn{-1_\delta}{1_{-\delta}}_B.
\end{equation}
Making the second $2\pi$ rotation we get $\nn{-1_\delta}{1_{-\delta}}_B \to \nn{3_\delta}{5_{-\delta}}_B$, so the asymptotics near
the point B are $\psi \sim \nn{3_\delta}{5_{-\delta}}_B$.  Now running over the big
semicircle to get back to the point A, as before, the WKB approximation allows us to claim the coefficient of the exponentially
big component $e^{-i \omega z}$ is unchanged.  So over the full round trip this component is multiplied by $\n{3_\delta}_B / \n{1}_A$.
Between \eqref{c1} and \eqref{c2} we have three homogeneous linear constraints on the four coefficients $A_\pm, B_\pm$, so we
can determine this ratio; it turns out to be
\begin{equation}
\frac{\n{3_\delta}_B}{\n{1}_A} = - \left(1 + 2\cos \pi j\right) - e^{2 i \omega \delta}(2+2\cos \pi j).
\end{equation}
The contour encloses only the singularity at $r=1$ (the outer horizon), so the monodromy comes only from the multivaluedness
of $\psi$ and $e^{-i \omega z}$ around $r=1$.  As discussed in the previous section, this monodromy is determined by the surface gravity
of the horizon at $r=1$, namely
\begin{equation}
\betaH = \frac{4\pi}{f'(1)} = \frac{4\pi}{1-k}.
\end{equation}
Also note that $2 i \delta = k^2 \beta$.  Then matching the monodromies we obtain the final result,
\begin{equation}
e^{\betaH \omega} = - \left(1 + 2\cos \pi j \right) - e^{k^2 \betaH \omega} \left(2 + 2\cos \pi j\right).
\end{equation}
The first term looks exactly like the result \eqref{higherf} from Schwarzschild, and then there is a
correction term which came from the phase shift $\delta$.  Actually, this term
can also be written in terms of a Hawking temperature, but this time the Hawking temperature of the inner horizon:
\begin{equation}
\betaI = \frac{4\pi}{f'(k)} = \frac{4\pi k^2}{k-1} = -k^2 \betaH.
\end{equation}
At any rate, substituting $j = 1/3$ or $j=5/3$, from \eqref{j-rn}, gives the same answer in each case:
\begin{equation} \label{rn-answer}
e^{\betaH \omega} = - 2 - 3e^{k^2 \betaH \omega}.
\end{equation}
So this is our result for the asymptotic quasinormal frequencies of scalar or axial electromagnetic-gravitational
perturbations of Reissner-Nordstr\o{}m in $d=4$.
This result suggests a somewhat more complicated asymptotic behavior for the quasinormal frequencies 
than we found in the case of Schwarzschild; in particular, the asymptotic frequencies seem to be
aperiodic unless $k^2$ is rational, and the real parts do not settle down to a simple value.  

We note that by taking $k$ slightly imaginary it
would be possible to get rid of the last term in the asymptotic limit, and in this case one would
obtain asymptotically $\abs{\re \omega} = \TH \log(2)$.  This may not be an unreasonable thing 
to do, since $k$ is related to the charge and mass of the black hole, and the mass
should be considered as slightly imaginary (the black hole eventually evaporates.)  Indeed, if $Q$
is real and fixed, then $M$ is a decreasing function of $k$, so the expected small negative imaginary
part of $M$ corresponds to a small positive imaginary part of $k$, which is the correct sign to 
suppress the last term.  Following \cite{dreyer,hod} and identifying this $\TH \log(2)$ with a quantum
of energy emitted by a black hole, we note that this corresponds precisely to the black hole reducing
its outer horizon area by $4 G \log(2)$ while keeping $Q$ fixed.

One might be concerned that our result for Reissner-Nordstr\o{}m is incompatible with the 
known results for Schwarzschild because 
it does not reduce to $\TH \log(3)$ as $k \to 0$.  There is no inconsistency;
rather, there is an order-of-limits issue.  For extremely small $k$, $\re \omega$
will be approximately $\TH \log(3)$ for a large range of $\im \omega$ but then eventually switch over to 
the Reissner-Nordstr\o{}m behavior determined by \eqref{rn-answer}; the value of $\im \omega$ at
which the crossover occurs goes to $\infty$ as $k \to 0$.
We can make a rough estimate of the critical value $k = k_0(\omega)$ 
above which the Reissner-Nordstr\o{}m description becomes more appropriate than the
Schwarzschild description; namely, it is the point
where the leading term $k^2/r^6$ is of the same order as the term
$1/r^4$ which would be leading for the Schwarzschild black hole.  To compare
these two terms the relevant $r$ is the place where we 
patch the plane wave with the Bessel function solution, namely $\omega z \approx 1$.
To relate $z$ to $r$ we can use either the relation
$z\sim r^2$ from Schwarzschild or $z \sim r^3/k$ from
Reissner-Nordstr\o{}m.  With either choice, requiring $k^2/r^6 \sim 1/r^4$, i.e. $k \sim r$, 
we obtain finally $k \sim r\sim \omega^{-1/2}$ i.e. 
\begin{equation}
k_0(\omega) \sim \omega^{-1/2}.
\end{equation}
If $k \gg k_0(\omega)$, the asymptotic Reissner-Nordstr\o{}m formula \eqref{rn-answer} 
is appropriate; if $k \ll k_0(\omega)$ the Schwarzschild result is appropriate.

After the preprint version of this paper 
appeared, the formula \eqref{rn-answer} as well as the order-of-limits issue 
discussed above were confirmed explicitly in the numerical analysis of \cite{bertikokkotas} (see also
\cite{neitzke} for some discussion.)

\section{Directions for further study}

This work suggests various possible directions for further study:

\begin{itemize}

\item The machinery developed here may be applicable more generally, e.g. to charged black holes in $d>4$,
and perhaps even to general rotating black holes.  Some work in this direction is in progress for the Kerr
black hole in $d=4$ \cite{inprog}.
In addition one could consider the case of black holes 
in de Sitter or anti-de Sitter space; quasinormal frequencies in these backgrounds have recently been
studied \cite{berti2,musiri}, and recently some exact results have been obtained in
the limit where the Schwarzschild horizon nearly coincides with the de Sitter horizon \cite{cardoso2}.
The asymptotic quasinormal frequencies in these backgrounds are related to properties of
the holographic dual theory \cite{horowitz, birmingham}.  The method will however require some
modification, since the potential no longer vanishes at infinity, and the expected answer is quite
different (the real part is expected to diverge along with the imaginary part \cite{berti2, horowitz}.)

\item Our results so far only give the asymptotic real part of the frequency in the limit of infinite
imaginary part.  It may be possible to extend the technique to obtain a systematic expansion
including corrections in $1 / \sqrt{\abs{\omega}}$.  After the preprint version of this paper appeared,
the $j$ and $l$ dependence of the first correction for arbitrary $j$ appeared in \cite{neitzke}, 
and the exact correction for $j=2$ including numerical coefficient was computed in \cite{mvb}.

\item It would be desirable to have some numerical calculations of the asymptotic quasinormal frequencies in
the cases we considered above, to confirm or reject our proposed analytic solution.  In the case of
Reissner-Nordstr\o{}m, after the preprint version of this paper appeared, such calculations were indeed
done \cite{bertikokkotas} and confirmed our solution in detail (the agreement is also discussed in \cite{neitzke}.)  However, the case of higher-dimensional black holes remains numerically unstudied as far as we know.

\item In $d>4$ Schwarzschild we could only give results for certain types of perturbation, because so far
a good theory of more general perturbations in these dimensions is apparently lacking.  It would be nice
to understand how the leading term in the potential depends on the type of perturbation.

\item According to our calculation, the asymptotic real part of the frequency depends 
only on the coefficient of the leading term in the potential.  
This leading term becomes important only near the singularity at $r=0$, but it still affects the
transmission amplitude in the physical region  between the horizon and infinity.  Although the
importance of the singularity at $r=0$ was purely mathematical in our
approach, there could be a more physical explanation of why our procedure
works.  Can we view it as a manifestation of the complementarity principle, reflecting some
kind of duality between what happens inside and outside the horizon?

\item The occurrence of the Hawking temperature for the inner horizon of the Reissner-Nordstr\o{}m
black hole is surprising.  How can the quasinormal modes know about the behavior of the black hole
in the causally disconnected interior?

\item The original motivation for this work was the LQG explanation \cite{dreyer} of the $\log(3)$ 
in $d=4$ Schwarzschild in terms of a spin network
with links of spin 1.  If LQG is correct, perhaps the more complicated
behavior for Reissner-Nordstr\o{}m should have a similar explanation.  The possible appearance
of $\log(2)$ in the asymptotic frequencies could support some recent claims \cite{corichi} that
the gauge group of LQG should be SU(2) despite the $\log(3)$ for Schwarzschild.

\item Similarly, can spin foam models or other models explain the fact that the $\log(3)$ for Schwarzschild 
black holes seems to be independent of dimension?

\end{itemize}

\noindent We have answered some questions about the asymptotic quasinormal frequencies
of black holes, but many more remain open.

\section*{Acknowledgements}

We are grateful to
N.\,Arkani-Hamed,
V.\,Cardoso,
K.\,Kokkotas,
K.\,Krasnov,
A.\,Maloney,
M.\,Schnabl,
and A.\,Stro\-min\-ger
for useful discussions, and to A. Maloney for reading a draft of this paper.
The work of L.M. was supported in part by
Harvard DOE grant DE-FG01-91ER40654 and the Harvard Society of Fellows.
The work of A.N. was supported by an NDSEG Graduate Fellowship.


\begin{thebibliography}{99}
%
%

\bibitem{dreyer}
O.~Dreyer,
{\it ``Quasinormal modes, the area spectrum, and black hole entropy,''}
[arXiv:gr-qc/0211076].

\bibitem{rovelli}
C.~Rovelli and P.~Upadhya,
{\it ``Loop quantum gravity and quanta of space: A primer,''}
[arXiv:gr-qc/9806079].

\bibitem{rovellitwo}
C.~Rovelli,
{\it ``Loop quantum gravity,''}
Living Rev.\ Rel.\  {\bf 1}, 1 (1998)
[arXiv:gr-qc/9710008].

\bibitem{thiemann}
T.~Thiemann,
{\it ``Introduction to modern canonical quantum general relativity,''}
[arXiv:gr-qc/0110034].

\bibitem{rovellismolin}
C.~Rovelli and L.~Smolin,
{\it ``Discreteness of area and volume in quantum gravity,''}
Nucl.\ Phys.\ B {\bf 442}, 593 (1995)
[Erratum-ibid.\ B {\bf 456}, 753 (1995)]
[arXiv:gr-qc/9411005].

\bibitem{ashle}
A.~Ashtekar and J.~Lewandowski,
{\it ``Quantum theory of geometry. I: Area operators,''}
Class.\ Quant.\ Grav.\  {\bf 14}, A55 (1997)
[arXiv:gr-qc/9602046].

\bibitem{barbero}
J.~F.~Barbero~G.,
{\it ``Real Ashtekar variables for Lorentzian signature space times,''}
Phys.\ Rev.\ D {\bf 51}, 5507 (1995)
[arXiv:gr-qc/9410014].

\bibitem{immirzi}
G.~Immirzi,
{\it ``Quantum gravity and Regge calculus,''}
Nucl.\ Phys.\ Proc.\ Suppl.\  {\bf 57}, 65 (1997)
[arXiv:gr-qc/9701052].

\bibitem{leaver}
E.W.\,Leaver,
{\it ``An analytic representation for the quasi-normal modes of Kerr black holes,''}
Proc.\ R.\ Soc.\ A {\bf 402} (1985) 285.

\bibitem{nollert}
H.-P.\,Nollert,
{\it ``Quasinormal modes of Schwarzschild black holes:
The determination of quasinormal frequencies 
with very large imaginary parts,''}
Phys.\ Rev.\ D {\bf 47} (1993) 5253-5258,
{\tt http://prd.aps.org/}.

\bibitem{andersson}
N.\,Andersson,
{\it ``On the asymptotic distribution of
quasinormal-mode frequencies for Schwarzschild black holes,''}
Class. Quantum Grav. {\bf 10}, L61-L67 (June 1993).

\bibitem{bachelot}
A.\,Bachelot, A.\,Motet-Bachelot,
{\it ``The resonances of a Schwarzschild black hole. (in French),''}
Annales Poincare Phys.\ Theor.\  {\bf 59}, 3 (1993).

\bibitem{hod}
S.~Hod,
{\it ``Bohr's correspondence principle and the area 
spectrum of quantum black holes,''}
Phys.\ Rev.\ Lett.\  {\bf 81}, 4293 (1998)
[arXiv:gr-qc/9812002].

\bibitem{lubosone}
L.\,Motl,
{\it ``An analytical computation of asymptotic\\
Schwarzschild quasinormal frequencies,''}
[arXiv:gr-qc/0212096].

\bibitem{glampand}
K.~Glampedakis and N.~Andersson,
{\it ``Quick and dirty methods for studying black-hole resonances,''}
[arXiv:gr-qc/0304030].

\bibitem{reggewheeler}
T.~Regge and J.~A.~Wheeler,
{\it ``Stability of a Schwarzschild singularity,''}
Phys.\ Rev.\ {\bf 108}, 1063 (1957).

\bibitem{andersson2}
N.~Andersson, K.~D.~Kokkotas and B.~F.~Schutz,
{\it ``A new numerical approach to the oscillation modes of relativistic stars,''}
Mon.\ Not.\ Roy.\ Astron.\ Soc.\  {\bf 274}, 1039 (1995)
[arXiv:gr-qc/9503014].

\bibitem{zerilli}
F.~J.~Zerilli,
{\it ``Gravitational field of a particle
falling in a Schwarzschild geometry analyzed in tensor harmonics,''}
Phys.\ Rev.\ D {\bf 2}, 2141 (1970).


\bibitem{ns}
H.-P.\,Nollert and B.~G.~Schmidt,
{\it ``Quasinormal modes of Schwarzschild black holes:  Defined
and calculated via Laplace transformation,''}
Phys.\ Rev.\ D {\bf 45} 2617 (1992).


\bibitem{kokkotas}
K.~D.~Kokkotas and B.~G.~Schmidt,
{\it ``Quasi-normal modes of stars and black holes,''}
Living Rev.\ Rel.\  {\bf 2}, 2 (1999)
[arXiv:gr-qc/9909058].

\bibitem{bd-horizon}
M.~E.~Araujo, D.~Nicholson and B.~F.~Schutz,
{\it ``On the Bohr-Sommerfeld formula for black hole normal modes,''}
Class.\ Quant.\ Grav.\ {\bf 10} 1127 (1993).

\bibitem{cardoso}
V.~Cardoso, O.~J.~Dias and J.~P.~Lemos,
{\it ``Gravitational radiation in $d$-dimensional spacetimes,''}
[arXiv:hep-th/0212168].

\bibitem{gibbons}
G.~Gibbons and S.~A.~Hartnoll,
{\it ``A gravitational instability in higher dimensions,''}
Phys.\ Rev.\ D {\bf 66}, 064024 (2002)
[arXiv:hep-th/0206202].

\bibitem{kunstatter}
G.~Kunstatter,
{\it ``d-dimensional black hole entropy spectrum from quasi-normal 
modes,''}
[arXiv:gr-qc/0212014].

\bibitem{bek2}
J.~D.~Bekenstein and V.~F.~Mukhanov,
{\it ``Spectroscopy of the quantum black hole,''}
Phys.\ Lett.\ B {\bf 360}, 7 (1995)
[arXiv:gr-qc/9505012].

\bibitem{bertikokkotas}
E.~Berti and K.~D.~Kokkotas,
{\it``Asymptotic quasinormal modes of Reissner-Nordstrom and Kerr black holes,''}
[arXiv:hep-th/0303029].

\bibitem{neitzke}
A.~Neitzke,
{\it ``Greybody factors at large imaginary frequencies,''}
[arXiv:hep-th/0304080].

\bibitem{inprog}
V.~Cardoso, L.~Motl and A.~Neitzke, work in progress.

\bibitem{berti2}
E.~Berti and K.~D.~Kokkotas,
{\it ``Quasinormal modes of Reissner-Nordstrom-anti-de Sitter black holes: scalar, electromagnetic and gravitational perturbations,''}
[arXiv:gr-qc/0301052].

\bibitem{musiri}
S.~Musiri and G.~Siopsis,
{\it `Quasinormal modes of large AdS black holes,''}
[arXiv:hep-th/0301081].

\bibitem{cardoso2}
V.~Cardoso and J.~P.~Lemos,
{\it ``Quasinormal modes of the near extremal Schwarzschild-de Sitter black hole,''}
[arXiv:gr-qc/0301078].

\bibitem{horowitz}
G.~T.~Horowitz and V.~E.~Hubeny,
{\it ``Quasinormal modes of AdS black holes and the approach to thermal  equilibrium,''}
Phys.\ Rev.\ D {\bf 62}, 024027 (2000)
[arXiv:hep-th/9909056].

\bibitem{birmingham}
D.~Birmingham, I.~Sachs and S.~N.~Solodukhin,
{\it ``Relaxation in conformal field theory, Hawking-Page transition, and  quasinormal/normal modes,''}
[arXiv:hep-th/0212308].

\bibitem{mvb}
A.~Maassen~van~den~Brink,
{\it ``WKB analysis of the Regge-Wheeler equation down in the frequency plane,''}
[arXiv:gr-qc/0303095].

\bibitem{corichi}
A.~Corichi,
{\it ``On quasinormal modes, black hole entropy, and quantum geometry,''}
[arXiv:gr-qc/0212126].



\end{thebibliography}
\end{document}